\documentclass[aps,pra,twocolumn]{revtex4}

\usepackage{amsmath,amsthm,amsfonts,xspace,verbatim}

\begin{document}
\newcommand{\ket}[1]{\left| #1 \right\rangle}
\newcommand{\bra}[1]{\left\langle #1 \right|}
\newcommand{\braket}[2]{\left\langle #1 | #2 \right\rangle}
\newcommand{\braopket}[3]{\bra{#1}#2\ket{#3}}
\newcommand{\proj}[1]{| #1\rangle\!\langle #1 |}
\newcommand{\Entropy}{H}
\newcommand{\KL}[2]{S\left(#1\|#2\right)}
\newcommand{\Tr}{\mathrm{Tr}}
\newcommand{\Rho}{P}
\def\Id{1\!\mathrm{l}}
\newcommand{\M}{\mathcal{M}}
\newcommand{\HSspace}{\textit{HS}-space\xspace}
\newcommand{\HSket}[1]{\left.\left| #1 \right\rangle\right\rangle}
\newcommand{\HSbra}[1]{ \left\langle\left\langle #1 \right.\right|}
\newcommand{\HSbraket}[2]{\left\langle\left\langle #1 | #2 \right\rangle\right\rangle}
\newcommand{\pspace}{\textit{p}-space\xspace}
\newcommand{\pket}[1]{\left| #1 \right)}
\newcommand{\pbra}[1]{\left( #1 \right|}
\newcommand{\pbraket}[2]{\left( #1 | #2 \right)}
\renewcommand{\log}{\ln}

\newtheorem{theorem}{Theorem}
\newtheorem{lemma}{Lemma}

\title{Accurate quantum state estimation via ``Keeping the experimentalist honest''}

\author{Robin Blume-Kohout}
\affiliation{Institute for Quantum Information, Caltech 107-81, Pasadena, CA 91125 USA}
\email{robin@blumekohout.com}
\author{Patrick Hayden}
\affiliation{School of Computer Science, McGill University,
        Montreal, QC H3A 2A7 Canada}
\email{patrick@cs.mcgill.ca}

\begin{abstract}
In this article, we derive a unique procedure for quantum state
estimation from a simple, self-evident principle:  an
experimentalist's estimate of the quantum state generated
by an apparatus should be constrained by honesty.  A skeptical
observer should subject the estimate to a test that guarantees
that a self-interested experimentalist will report the true state
as accurately as possible. We also find a non-asymptotic, operational
interpretation of the quantum relative entropy function.
\end{abstract}

\maketitle

Consider a source of quantum states such as a laser, or an ion trap
with a preparation procedure.  Quantum state estimation is the
problem of deducing \emph{what} state it emits by analyzing the
outcomes of measurements on many instances. The usual procedure for
state estimation is \emph{quantum state tomography}~\cite{VogelPRA89,
DArianoAIEP03}, together with some variant of \emph{maximum-likelihood
estimation} \cite{HradilPRA97,HradilLNP04} to ensure positivity.  The
obvious goal is an estimate ``close'' to the true state.
Different metrics, such as fidelity~\cite{BaganPRA05},
relative entropy~\cite{TanakaPRA05}, trace norm, or Hilbert-Schmidt
norm~\cite{Rehacek04}, will favor different estimation procedures.

Here, we derive an optimal state-estimation procedure by first
identifying quantum relative entropy as a unique metric for characterizing
an estimate's ``goodness''. Our procedure is broadly
adaptable to (1) arbitrary prior knowledge (or ignorance)
and (2) arbitrary measurement procedures.



\textbf{Keeping the Experimentalist Honest: }
Implicit in the idea of state estimation is the assumption that
some estimates are better than others.  Suppose that $\sigma$ is
an estimate of the true state $\rho$, and that $f(\rho:\sigma)$
is a measure of how ``good'' an estimate $\sigma$ is.  We propose
that this measure should obey three principles:
\begin{enumerate}
 \item The best estimate of $\rho$ is $\rho$ itself.  If
$f(\rho:\sigma)$ measures how well $\sigma$ estimates $\rho$, then
$f(\rho:\rho) > f(\rho:\sigma)$ for all $\sigma\neq\rho$.
 \item $f(\rho:\sigma)$ should correspond to some \emph{operational}
test, as the payoff or cost of some experimental procedure.
 \item The ``reward'' for correctly predicting an event
should depend only on the predicted probability for \emph{that}
event.  This is a version of the likelihood principle (see
\cite{BernardoAnnStat79}).
\end{enumerate}
Remarkably, these simple assumptions single out one measure: the
\emph{relative entropy}
between $\rho$ and $\sigma$, or $\KL{\rho}{\sigma} =
\Tr(\rho\ln\rho-\rho\ln\sigma)$.  It arises as the expected payoff
in a type of game between a cash-strapped experimentalist and her employer.

Alice, an ambitious scientist attempting to build a quantum
computer, produces states that she believes are described by
the density operator $\rho$.  She informs her employer, Bob,
that she has produced the state $\sigma$.  Bob, a conscientious
scientific administrator, would like to ensure that Alice does not
lie -- that $\sigma = \rho$.  He will periodically visit
Alice's lab and measure one of her states, in a way that may depend
on her estimate $\sigma$.  Her future funding will depend on the
outcomes of these measurements. What measurement should Bob perform,
and how should he pay Alice, so that she has no incentive to deceive him?

We propose that Bob should measure in a basis $\{\ket{f_i}\}_{i=1}^n$
that diagonalizes $\sigma = \sum_i s_i \proj{f_i}$.  Upon getting
outcome $i$, he should pay Alice $R_i = C + D \log s_i$ dollars, where
$C$ and $D$ are non-negative constants.  We denote this as the ``honest
experimentalist reward scheme,'' or HERS.

\textbf{HERS motivates honesty}:  Bob's measurement yields outcome $i$
with probability $p_i = {\rm Tr}( \rho
\proj{f_i})$.  Alice's expected reward is
\begin{equation}
\overline{R} = \sum_{i=1}^n{p_i R_i} = C + D \sum_{i=1}^n{ p_i \log s_i}.
\end{equation}
Rewriting the last term as $\sum_i{ p_i \log s_i} = \Tr\rho\log\sigma$ yields
\begin{eqnarray}
\overline{R}(\rho:\sigma) &=& C + D \, \Tr\rho\log\sigma \nonumber \\
&=& C + D\left[ \Tr\rho\log\rho - (\Tr\rho\log\rho - \Tr\rho\log\sigma) \right] \nonumber \\
&=& C - D\left[ \Entropy(\rho) + \KL{\rho}{\sigma} \right]
\end{eqnarray}
Since $C$, $D$, and $\rho$ are fixed, Alice maximizes her expected
reward by reporting a $\sigma$ that minimizes the relative entropy
$\KL{\rho}{\sigma}$.  This constrains $\sigma$ to be $\rho$
itself~\cite{OhyaBook93}. Alice is thereby motivated to be honest.
She is also motivated to produce pure states -- but \emph{not} to
lie about how pure the actual state is.

\textbf{HERS is unique}:  Unless Bob can do non-projective POVMs
\footnote{Characterizing Bob's \emph{non}-projective
measurements is more complicated.  For now, we note that: (1)
in general, a fixed informationally complete POVM suffices;
(2) our results remain largely unchanged, but Alice's expected reward
may involve $\mathcal{C}[ \rho ]$ and $\mathcal{C}[\sigma ]$ instead of
$\rho,\sigma$, where $\mathcal{C}$ is a quantum channel.},
this turns out to be the \emph{only} verification procedure that
satisfies our three criteria.

In classical statistics, a reward scheme for a probabilistic
forecast is a \emph{scoring rule}.  It assigns an average
reward $\overline{R}( P : Q )$ to a forecast $Q$ when events
are distributed according to $P$.  A reward that is uniquely
maximized by an honest forecast is a \emph{strictly proper
scoring rule} or SPSR (see review \cite{Gneiting05}).
Given some $P$, the maximum reward under such a rule
is $P$'s \emph{value}, $G(P) \equiv \overline{R}( P : P )$.
Savage showed that for every SPSR, $G(P)$
is strictly convex~\cite{SavageJASA71}.

To consider the quantum case, we observe that a measurement transforms
a state $\rho$ into a probability distribution $\{P_i\}$ over outcomes,
to which we can apply a scoring rule.  We represent a projective
measurement of basis $B$ as a quantum channel $\beta$.  For any state $\rho$, let
$\beta[\rho] = P$.  $P$ is a diagonal matrix of probabilities, and
$\beta$ simply annihilates off-diagonal elements.  Let $G(\rho)$ be
the maximum of $G(\beta[\rho])$ over all $\beta$.
Since the eigenvalues of $\rho$ majorize those of $\beta[\rho]$
\cite{BhatiaBook96}, and $G$ is convex, this maximum is achieved
when $\beta[\rho] = \rho$.  Thus $G(\rho) = G(\{\lambda_i\})$, where
$\{\lambda_i\}$ are the eigenvalues of $\rho$.

\begin{lemma}
\label{LemmaMeasurement} Given a physical state $\rho$ and an
estimate $\sigma$, Bob can ensure Alice's honesty by applying a
SPSR to the probabilities for a measurement
of basis $B$ if and \textbf{only} if $\sigma$ is diagonal in $B$.
\end{lemma}

\textbf{Proof}:  Represent Bob's $\sigma$-dependent measurement
as a CP-map $\beta_\sigma$ that annihilates off-diagonal
elements in basis $B$.  Let the SPSR yield a value $G$.

1.  Suppose that $B$ diagonalizes $\sigma$, so
$\beta_\sigma[\sigma] = \sigma$.  Then Alice's expected reward is
\begin{equation}
\overline{R}( \beta_\sigma[\rho] : \beta_\sigma[\sigma] ) \leq G(\beta_\sigma[\rho]) \leq G(\rho).
\end{equation}
The inequalities are simultaneously saturated if and only
if $\sigma = \rho$, in which case Alice earns the full value $G(\rho)$
of her state.  When $\sigma \neq \rho$, one or both of the inequalities is strict,
so Alice earns strictly less than $G(\rho)$.  Thus, Alice maximizes her reward uniquely
by reporting $\sigma=\rho$.

2.  Suppose that there exists a $\sigma$ so that $\beta_\sigma[ \sigma ]
\neq \sigma$.  Then let $\rho = \beta_\sigma[ \sigma ]$.
Since $\beta_\sigma^2 = \beta_\sigma$, $\beta_\sigma[ \rho ] =
\rho = \beta_\sigma[ \sigma ]$.  Alice's expected reward is
\begin{equation}
\overline{R}( \beta_\sigma[ \rho ] : \beta_\sigma[ \sigma ] ) = \overline{R}( \rho, \rho) = G( \rho ).
\end{equation}
whereas if she (truthfully) reports $\rho$, she can expect
\begin{equation}
\overline{R}( \beta_\rho[ \rho ] : \beta_\rho[ \rho ] ) = G( \beta_\rho[\rho] ) \leq G( \rho ),
\end{equation}
where the inequality holds because $\rho$'s eigenvalues majorize
those of $\beta_\rho[ \rho ]$ (by Schur's theorem~\cite{BhatiaBook96}),
and because $G$ is convex.  Alice expects the same reward for reporting
$\rho$ or $\sigma\neq\rho$, so her honesty is not ensured. \qed

So far we have demanded only that our scoring rule be strictly proper.
We now demand that Alice's reward depend only on her predicted probability
for the \emph{observed} event. In other words, $R_i(\{s_j\}) = R_i(s_i)$.
This reflects the Likelihood Principle \cite{Berger88}: all the
relevant information in an event is contained in the likelihood of the
hypothesis (here, $p(i | \sigma)$).  \emph{How} the experiment was
performed is irrelevant.  In particular, this avoids any argument between
Alice and Bob about how to describe the outcome[s] that did not occur.

A remarkable theorem by Aczel (see also \cite{BernardoAnnStat79}) then
restricts the form of the reward function $R_i( s_i )$.

\begin{theorem}[Aczel \cite{AczelJMAA80}]
\label{thm:Aczel}
Let $n\geq 3$.  The inequality
\begin{equation}
\sum_{i=1}^n p_i R_i(q_i) \leq \sum_{i=1}^n p_i R_i(p_i)
\end{equation}
is satisfied for all $n$-point probability distributions
$(p_1\ldots p_n)$ and $(q_1\ldots q_n)$ if and only if there
exist constants $C_1\ldots C_n$ and $D$ such that for all $i\in[1\ldots n]$,
\begin{equation}
R_i(p) = D \log p + C_i.
\end{equation}
\end{theorem}

In the scenario we consider, there is even less freedom.  Aczel's theorem
allows the constants $C_i$ to depend on $i$.  In the quantum setting, all
the $C_i$ must be equal to a fixed $C$ \emph{independent} of $i$ (see proof
in Appendix \ref{app:Ci_equals_C}). We have therefore proved the following:
\begin{theorem}[The honest experimentalist]
\label{thm:honest_experimentalist}
Let $A$ be a quantum system with dimension $n\geq 3$. Let $\rho$ and $\sigma$
be density operators for $A$, and let $\{\ket{g_i}\}$ be an orthonormal basis
for $A$ that depends only on $\sigma$.  Defining $p_i = \bra{g_i}\rho\ket{g_i}$ and $s_i =
\bra{g_i}\sigma\ket{g_i}$, suppose that
\begin{equation}
\sum_i p_i R_i ( s_i ) \leq \sum_i p_i R_i(p_i)
\end{equation}
is satisfied for all $\rho$ and $\sigma$, with equality if and only if
$\rho = \sigma$.  Then $\{\ket{g_i}\}$ diagonalizes
$\sigma$, and there exist constants $C$ and $D$ such that $R_i( s_i
) = C + D \ln s_i$ for all $i$.
\end{theorem}

The scheme outlined (HERS) is \emph{uniquely} specified by Bob's need
to guarantee Alice's honesty through self-interest.  Relative entropy,
$\KL{\rho}{\sigma}$, appears naturally as the amount of money that
Alice can expect to lose by lying.  Any ``boss'' who wishes never to
be lied to must use the HERS payment scheme.

Of course, rewards in the real world are often structured less wisely.
We propose, however, that a maximally ethical scientist should
\emph{act} as if she were being motivated by HERS, and use $\KL{\rho}{\sigma}$ as
the universal measure of \emph{honesty}.  Hereafter, we will assume that
the experimentalist is, in fact, honest.



\textbf{The Uncertain Experimentalist: }
What does ``honesty'' mean for an experimentalist who is not certain of $\rho$?
We assert that she should behave \emph{as if} she were guided by a
tangible, strictly proper, reward scheme.  HERS is an excellent candidate,
but our proofs hold for any strictly proper scheme.


Suppose that Alice does not know $\rho$, but knows that it will be selected
from an ensemble $\pi(\rho)\mathrm{d}\rho$ (or simply $\pi(\rho)$ hereafter, for clarity).  Equivalently, she thinks the true state
is $\rho$ with probability $\pi(\rho)$.  Her expected reward (from HERS)
for reporting $\sigma$ is:
\begin{eqnarray}
\overline{R} &=& \int{\!R(\rho:\sigma)\pi(\rho)\mathrm{d}\rho} \nonumber \\
&=& C - D\left(\int{\!\Entropy(\rho)\pi(\rho)\mathrm{d}\rho} + \int{\!\KL{\rho}{\sigma}}\pi(\rho)\mathrm{d}\rho\right) \nonumber \\
&=& \mathrm{const}_\sigma + D\int{\!\Tr(\rho\log\sigma)\pi(\rho)\mathrm{d}\rho} \nonumber \\
&=& \mathrm{const}_\sigma + D\left[\Tr(\overline{\rho}\log\sigma)\right]
= \mathrm{const'}_\sigma - D\left[\KL{\overline{\rho}}{\sigma}\right], \nonumber
\end{eqnarray}
where $\overline{\rho} = \int{\rho\pi(\rho)\mathrm{d}\rho}$.
The ``$\mathrm{const}_\sigma$'' terms are independent of $\sigma$
and therefore out of Alice's control.  Therefore, Alice maximizes
her honesty by reporting the mean of her probability distribution.

The uniqueness of HERS depends on the Likelihood Principle.
However, the mean of the probability distribution is maximally honest
for \emph{any} strictly proper scoring rule:

\begin{theorem}
\label{thm:mean}
Let Alice believe that $\rho$ is selected from a distribution $\pi(\rho)$.
Let her expected reward for reporting $\sigma$ be
$R(\rho:\sigma) = \sum_i{p_i R_i(\sigma)}$, where $p_i = \Tr E_i\rho$, and $R(\rho:\rho) > R(\rho:\sigma)$ for all $\sigma \neq \rho$.
Alice maximizes her expected reward by reporting
$\sigma = \overline{\rho} \equiv \int{\rho\pi(\rho)\mathrm{d}\rho}$.
\end{theorem}

\textbf{Proof:} Since Alice expects $\rho$ to appear with probability
$\pi(\rho)$, her expected reward is:
\begin{eqnarray}
\overline{R} &=& \int{R(\rho:\sigma)\pi(\rho)\mathrm{d}\rho} \\
 &=& \int{ \sum_i{\Tr (E_i\rho) R_i(\sigma)}\pi(\rho)\mathrm{d}\rho} \\
 &=& \sum_i{\Tr E_i\left(\int{\rho\pi(\rho)\mathrm{d}\rho}\right) R_i(\sigma) } \\
 &=& R( \overline{\rho}:\sigma ).
\end{eqnarray}
$R$ is strictly proper, so the unique maximum of $R( \overline{\rho}:\sigma )$
is at $\sigma = \overline{\rho}$. \qed

Consider, instead if Alice had tried to maximize fidelity \cite{JozsaJMO94},
which is not derived from an operational procedure,
but \emph{would} guarantee Alice's honesty when she knows $\rho$
exactly~\footnote{Jozsa showed \cite{JozsaJMO94} that $F(\sigma,\rho)$
is maximal if and only if $\sigma=\rho$}.
Suppose that Alice knows that $\rho$ is either $\proj{0}$ or
$\proj{+}$, with equal probability.
The fidelity between any $\sigma$ and a pure state is
$F(\sigma,\proj{\psi}) = \braopket{\psi}{\sigma}{\psi}$,
so the \emph{average} fidelity is just:
\begin{equation}
\overline{F} = \Tr\left(\sigma\overline{\rho}\right),
\end{equation}
where $\overline{\rho} = \frac12\left(\proj{0}+\proj{+}\right)$.
To maximize $\overline{F}$, Alice would choose the largest
eigenstate of $\overline{\rho}$ -- \emph{not} $\overline{\rho}$
itself.  Thus, while fidelity appears at first like a good measure of honesty,
it does \emph{not} generally motivate Alice to report the mean of
her distribution.  Moreoever, it can motivate her to report a pure state
that she knows is not the true state.

This is not simply a different definition of honesty.  An
experimentalist who reports a pure state $\ket{\psi}$ is
predicting that some event will \emph{never} occur.  If
Alice reports $\ket{0}$, she is asserting that no measurement
will ever yield $\ket{1}$. In the presence of any uncertainty
whatsoever, this is at best misleading, and at worst an outright lie.

This illustrates that HERS strongly penalizes over-optimism.  If Bob
obtains an outcome $\proj{f_i}$ for which $\braopket{f_i}{\sigma}{f_i}=0$,
Alice will lose infinitely much money!  A truly zero-probability
event is one against which a gambler would bet infinite money, at
arbitrarily bad odds.  Reporting $p=0$ for an event that could conceivably
happen is infinitely misleading, and should be discouraged.


\textbf{The Informed Experimentalist: }
How should Alice use the results of measurements (that she has performed)
to reduce her uncertainty?
Suppose that she has performed POVM measurements on $N$ copies of $\rho$, where
the $i$th result corresponds to a positive operator $E_i$.  She knows
two things:
\begin{enumerate}
\item $\rho$ is selected at random from an ensemble
described by $\pi_0(\rho)$.
\item Through experiments on copies of $\rho$, she has obtained a
measurement record $\M=\left\{E_1,E_2\ldots E_N\right\}$.
\end{enumerate}


Suppose that she reports $\sigma_j$ when $\M_j$
occurs, and is paid according to a SPSR where $\overline{R}(\rho:\sigma) = \sum_i{\Tr(\proj{f_i}\rho)R_i(\sigma)}$.

Since $\rho$ appears with probability $\pi_0(\rho)$, the event ``$\rho$
appeared, $\M_j$ was measured, and $\sigma_j$ was reported'' occurs with
probability $\pi_0(\rho)p\left(\M_j|\rho\right)$.  Alice's expected reward
over \emph{all} possible events is:
\begin{eqnarray}
\overline{R} &=& \int{ \sum_j{ R(\rho:\sigma_j) p\left(\M_j|\rho\right) \pi_0(\rho) \mathrm{d}\rho}} \nonumber \\
&=& \sum_j{\int{ \left(\sum_i{\Tr(\proj{f_i}\rho) R_i(\sigma_j)}\right) p\left(\M_j|\rho\right) \pi_0(\rho)\mathrm{d}\rho}} \nonumber \\
&=& \sum_{i,j}{\Tr\left[\proj{f_i}\left(\int{\rho p\left(\M_j|\rho\right) \pi_0(\rho) \mathrm{d}\rho} \right)\right] R_i(\sigma_j)}.\nonumber
\end{eqnarray}
We rewrite this using
\begin{eqnarray}
p_j &\equiv& \int{p\left(\M_j|\rho\right) \pi_0(\rho) \mathrm{d}\rho}\mbox{, and} \\
\overline{\rho}_j &\equiv& \frac{1}{p_j}\int{\rho p\left(\M_j|\rho\right)\pi_0(\rho) \mathrm{d}\rho} \label{EqBayes},
\end{eqnarray}
to get
\begin{equation}
\overline{R} = \sum_j{p_j\Tr\left(\proj{f_i}\overline{\rho}_j\right)R_i(\sigma_j)} = \sum_j{p_j R(\overline{\rho}_j:\sigma_j)}.\nonumber
\end{equation}
$\overline{R}$ is strictly proper, so by setting
$\sigma_j = \overline{\rho}_j$ we uniquely maximize each term in
the sum, and Equation \ref{EqBayes} defines the optimal estimate
of $\rho$, given $\M_j$.

Equation \ref{EqBayes} is nothing other than Bayes' Rule.  Thus, the mean
of a Bayesian-inferred distribution over states is the unique optimal
estimate -- for \emph{any} strictly proper reward scheme.
We formalize this in the following theorem:

\begin{theorem}
If $\rho$ is drawn from an ensemble $\pi_0(\rho)$, and a measurement $\M_j$
with conditional probability $p\left(\M_j|\rho\right)$ is observed, then
every strictly proper scoring rule $\overline{R}(\rho:\sigma)$ is maximized
by:
\begin{enumerate}
\item Using Bayes' Rule and Born's Rule:
\begin{eqnarray}
\pi_0(\rho) \longrightarrow \pi_\M(\rho) &=& \frac{p(\M|\rho)\pi_0(\rho)}{\int{\mathrm{d}\rho\pi_0(\rho)p(\M|\rho)}} \nonumber\\
&=&\frac{\left[\prod_{i=1}^N{\Tr(E_i\rho)}\right]\pi_0(\rho)}{\int{\mathrm{d}\rho\left[\prod_{i=1}^N{\Tr(E_i\rho)\pi_0(\rho)}\right]}}.\nonumber
\end{eqnarray}
\item Reporting the mean of $\pi_\M(\rho)$.
\end{enumerate}
\end{theorem}

This applies not only to relative entropy, our preferred measure of
honesty, but to \emph{any} honesty-guaranteeing reward scheme.
We conclude that Bayesian inference is the unique solution to
\emph{honest} state estimation.

Other procedures will not optimize any measure of honesty derived from a
strictly proper scoring rule.  Alternative measures of honesty will either
(a) in some circumstances, motivate an experimentalist to flat-out lie
about the state, or (b) not be operationally implementable (e.g., fidelity).
Our previous discussion of fidelity illustrates that a non-operational metric
that guarantees the honesty of a \emph{knowledgeable} experimentalist can fail
dramatically in the face of uncertainty.



Information theorists have previously interpreted relative entropy
as a good measure of two states' distinguishability \cite{VedralPRA97,
FuchsThesis95} -- indeed, as the only meaningful one in the limit of
many copies.  We have invoked the Likelihood Principle rather than
the many-copy limit, but we can easily allow Bob to jointly measure $N$
copies of $\rho$.  He must then apply a SPSR to the result, and Alice can
expect a reward $\overline{R}(\rho^{\otimes N}:\sigma^{\otimes N})$.
As $N\rightarrow\infty$, relative entropy remains meaningful, unlike
other measures (e.g., the Brier score \cite{Gneiting05}).

In the presence of uncertainty, pure (or rank-deficient)
states are infinitely dishonest estimates.  Estimating a pure state
means predicting that some event will \emph{never} happen.  Anyone
taking such a prediction seriously would be justified in betting
infinitely much money, at arbitrarily bad odds, against that event
-- and should therefore expect to lose infinitely much, if the
estimate is incorrect.  This should be discouraged.

Bayesian state estimation has been discussed previously
\cite{DerkaJFMO96,SchackPRA01,Neri05}, especially for pure states
\cite{JonesAOP91,JonesPRA94}.  The predominance of other methods
such as maximum likelihoood, in the current literature (e.g., \cite{JamesPRA01,ReschPRL05,WeinsteinJCP04,OBrienPRL04,MyrskogPRA05,
RoosPRL04,AltepeterPRL03,HaeffnerNature05}, indicates that it has
not received the attention it deserves.  Our goal in this letter is
to provide a concrete and compelling argument for Bayesian state
estimation -- and to call attention to the problematic implications of
pure-state estimates.

\begin{acknowledgments}
RBK thanks Daniel James (for pointing out the problem in the first
place), Jon Yard, John Preskill, and NSF contract PHY-0456720. PH
thanks Howard Barnum, Simon Benjamin and especially Chris Fuchs
for helpful discussions, as well as the Canada Research Chairs
program, the Canadian Institute for Advanced Research and NSERC
for support.
\end{acknowledgments}

\appendix

\section{Further restricting the reward function}
\label{app:Ci_equals_C}

We show that the quantum reward function is more tightly constrained
than the classical one: the constants $C_i$ of Theorem \ref{thm:Aczel}
must all be equal. We can assume that $D=1$.  Assuming a
reward function of the form specified by Theorem
\ref{thm:Aczel}, the inequality $\overline{R}(\rho:\sigma) \leq
\overline{R}(\rho:\rho)$ is then equivalent to
\begin{equation} \label{eqn:Ci_elimination}
\KL{\rho}{\sigma} + \sum_i C_i ( r_i - p_i ) \geq 0.
\end{equation}
Let $U(t)$ be a smooth curve in the unitary group defined on a
neighborhood of $t=0$ and such that $U(0) = I$. Also let
$\sigma(t) = U(t) \rho U^\dagger(t)$ and
\begin{equation}
g(t) = \KL{\rho}{\sigma(t)} + \sum_i C_i \big[ r_i - \Tr\big(\rho
U(t)\proj{e_i}U^\dagger(t) \big) \big]\nonumber
\end{equation}
be the function defined by substituting $\sigma(t)$ into the
expression of Eq.~(\ref{eqn:Ci_elimination}). (Here $\rho = \sum_i
r_i \proj{e_i}$, so  that $\ket{f_i} = U \ket{e_i}$.)
Differentiating gives
\begin{equation}
\frac{d}{dt}\KL{\rho}{\sigma(t)}
 = -\Tr\big[ \rho \big(
    \dot{U}(\ln \rho)U^\dagger + U (\ln \rho) \dot{U}^\dagger
    \big) \big],
\end{equation}
where $\dot{U} = dU/dt$ and the dependence of $U$ on $t$ has been
suppressed. Because $\dot{U} + \dot{U}^\dagger = 0$ when $t=0$,
$\mbox{$\frac{d}{dt}$}\KL{\rho}{\sigma(t)}\vert_{t=0} = 0$.
Likewise, $\dot{p}_i$ equals $0$ when $t=0$.
We must therefore determine the second derivative of $g$ at $t=0$
and show that for a suitable choice of curve, this derivative is
negative. In that case, $g(t) = g(0) + \ddot{g}(0) t^2 /2 +
O(t^3)$ with $\ddot{g}(0) < 0$, implying that $g(t)$ is negative
for sufficiently small $t$.

So, differentiating again, we find
\begin{eqnarray}
\frac{d^2}{dt^2} \KL{\rho}{\sigma(t)} \Big\vert_{t=0}
 &=& 2 \, \Tr\Big[ [X,\rho] (\ln \rho) X \Big], \label{eqn:ddKL}
\end{eqnarray}
where $X$ is a Hermitian matrix such that $\dot{U} = iX$. We have
made use of the identity $\ddot{U} + \ddot{U}^\dagger = - 2
\dot{U}\dot{U}^\dagger$ that can be proved by differentiating
$UU^\dagger = I$. Because $\KL{\rho}{\sigma(t)} \geq 0$, the
expression in Eq.~(\ref{eqn:ddKL}) must also be nonnegative.

Differentiating the second term of $g(t)$, we find that
\begin{equation} \label{eqn:ddg}
\frac{d^2 g}{dt^2}\Big\vert_{t=0}
 = 2 \, \Tr\Big[ [X,\rho](\ln \rho + B)X\Big],
\end{equation}
where $B = \sum_i C_i \proj{e_i}$.

We will now show that all the $C_i$ must be equal. Assume without
loss of generality that $C_1 \neq C_2$. Let $r_j
= \exp(-(C_j+2\ln 2)/2)$ for $j=1,2$ and $r_3
= 1 - r_1 - r_2$. With these choices,
$r_1, r_2 \leq 1/2$, making $\rho$ a density operator. Now write
$\tilde{\rho}$  and $\tilde{B}$ for the restriction of $\rho$ and
$B$ to the span of the first two eigenvectors of $\rho$. Observe
that there exists a choice of $X$ also with support only on this
subspace such that $\Tr\big[ [X,\tilde{\rho}] (\ln \tilde{\rho}) X
\big] > 0$.

With these choices, and noting that $\Tr\big[
[X,\tilde{\rho}]X\big] = 0$,
\begin{eqnarray}
\ddot{g}(0)
 &=& -2 \, \Tr\Big[ [X,\tilde{\rho}](\ln \tilde{\rho})X \Big] < 0.
\end{eqnarray}
Because $\ddot{g}(0) < 0$ contradicts the
requirement that $D(\rho\|\sigma(t)) \geq 0$, we conclude that
$C_i = C_1$ for all $i$. \qed

\bibliographystyle{apsrev}
\bibliography{quantum,Estimation,math}

\end{document}